\documentclass{PoS}

\usepackage{bm}
\usepackage{amsmath, amssymb}

\title{Theoretical analysis on the $K^{-} {}^{3} \text{He} \to \Lambda
  p n$ reaction for the $\bar{K} N N$ bound-state search in the J-PARC
  E15 experiment}

\ShortTitle{Theoretical analysis on the $K^{-} {}^{3} \text{He} \to
  \Lambda p n$ reaction}

\author{\speaker{Takayasu Sekihara}\\
  Advanced Science Research Center, Japan Atomic Energy Agency,
  Shirakata, Tokai, Ibaraki, 319-1195, Japan\\
  E-mail: \email{sekihara@post.j-parc.jp}}

\author{Eulogio Oset\\
  Departamento de F\'{\i}sica Te\'orica and IFIC, Centro
  Mixto Universidad de Valencia-CSIC, Institutos de Investigaci\'on
  de Paterna, Aptdo. 22085, 46071 Valencia, Spain\\
  E-mail: \email{oset@ific.uv.es}}

\author{Angels Ramos\\
  Departament de F\'{\i}sica Qu\`antica i Astrof\'{\i}sica
  and Institut de Ci\`encies del Cosmos, Universitat de Barcelona,
  Mart\'i i Franqu\`es 1, 08028 Barcelona, Spain\\
  E-mail: \email{ramos@fqa.ub.edu}}

\abstract{We theoretically analyze the $K^{-} {}^{3} \text{He} \to
  \Lambda p n$ reaction for the $\bar{K} N N$ bound-state search in
  the J-PARC E15 experiment.  We find that, by detecting a fast and
  forward neutron in the final state, an almost on-shell $\bar{K}$ is
  guaranteed, which is essential to make a bound state with two
  nucleons from ${}^{3} \text{He}$.  Then, this almost on-shell
  $\bar{K}$ can bring a signal of the $\bar{K} N N$ bound state in the
  $\Lambda p$ invariant-mass spectrum, although it inevitably brings a
  kinematic peak above the $\bar{K} N N$ threshold as well.  As a
  consequence, we predict two peaks across the $\bar{K} N N$ threshold
  in the spectrum: the lower peak coming from the $\bar{K} N N$ bound
  state, and the higher one originating from the kinematics. }

\FullConference{XVII International Conference on Hadron Spectroscopy
  and Structure - Hadron2017\\ 25-29 September, 2017\\ University of
  Salamanca, Salamanca, Spain}

\begin{document}

\section{Introduction}

Among various combinations of hadrons, the antikaon ($\bar{K}$) and
nucleon ($N$) form one of the most interesting pairs.  The $\bar{K}$
meson is a Nambu-Goldstone boson of the spontaneous chiral symmetry
breaking of quantum chromodynamics (QCD), which constrains the
$\bar{K} N$ interaction to be strongly attractive in a model
independent manner.  This chiral $\bar{K} N$ interaction together with
its coupled channels dynamically generates the $\Lambda (1405)$
resonance~\cite{Kaiser:1995eg, Oset:1997it, Oller:2000fj, Oset:2001cn,
  Lutz:2001yb, Jido:2003cb}.  Recently it was shown that the $\Lambda
(1405)$ resonance in chiral dynamics was indeed a $\bar{K} N$ bound
state~\cite{Sekihara:2014kya, Kamiya:2015aea} in terms of the
compositeness~\cite{Hyodo:2011qc, Aceti:2012dd, Sekihara:2016xnq}.

Because the $\bar{K} N$ interaction is attractive enough to make a
bound state as the $\Lambda (1405)$ resonance, we expect that there
should exist bound states of $\bar{K}$ and nuclei, which are called
kaonic nuclei.  Motivations to study kaonic nuclei are: they are
exotic states of many-body systems interacting strongly, and provide
us with a test field of the $\bar{K} N$ interaction and behavior of a
strange quark in finite nuclear density.  For kaonic nuclei, in
particular for the simplest kaonic nucleus, \textit{i.e.}, the
$\bar{K} N N$ bound state or the ``$K^{-} p p$'' state, many
experimental searches and theoretical predictions have been performed,
but even their existence is still controversial (see review in
Ref.~\cite{Nagae:2016cbm}).

\begin{figure}[b]
  \centering
  \includegraphics[width=8.6cm]{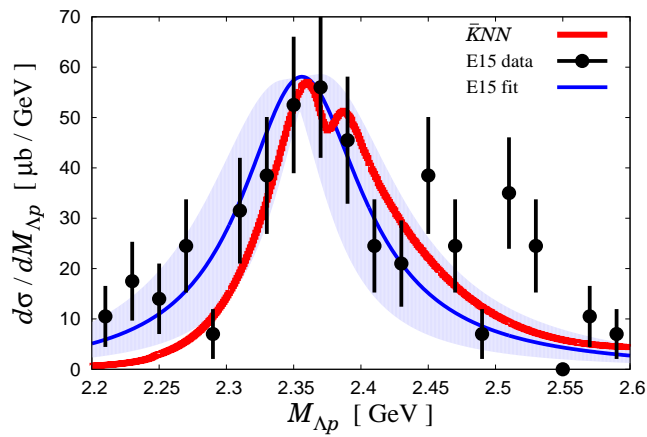}
  \caption{$\Lambda p$ invariant-mass spectrum of the $K^{-}
    {}^{3}{\rm He} \to \Lambda p n$ reaction~\cite{Sekihara:2017ncl}.
    Our theoretical result, shown as the thick red line, is obtained
    in the scenario that the $\bar{K} N N$ bound state is
    generated~\cite{Sekihara:2016vyd}.  The experimental (E15) data
    and its fit are taken from Ref.~\cite{Sada:2016nkb} and shown in
    arbitrary units.}
  \label{fig:dsdM}
\end{figure}

In this line, the result from the J-PARC E15
experiment~\cite{Hashimoto:2014cri, Sada:2016nkb} is very promising.
In the J-PARC E15 experiment, they observed the $K^{-} {}^{3}
\text{He} \to \Lambda p n$ reaction with the initial kaon of momentum
$1 \text{ GeV} / c$, a fast and forward neutron in the final state,
and no spectator nucleon.  As a result of the E15 first run, they
found a peak structure near the $K^{-} p p$ threshold as the black
points and blue bands in Fig.~\ref{fig:dsdM}, which could be a signal
of a $\bar{K} N N$ bound state.  In order to understand the reaction
mechanism and to investigate how this peak is constructed, we perform
a theoretical analysis on this reaction.  Details of the calculations
are provided in Refs.~\cite{Sekihara:2016vyd, Sekihara:2016gjh,
  Sekihara:2017ncl}.

\section{Theoretical analysis on the $K^{-} {}^{3} \text{He} \to
  \Lambda p n$ reaction}

\begin{figure}[t]
  \centering
  \begin{tabular}{cc}
    \includegraphics[scale=0.16]{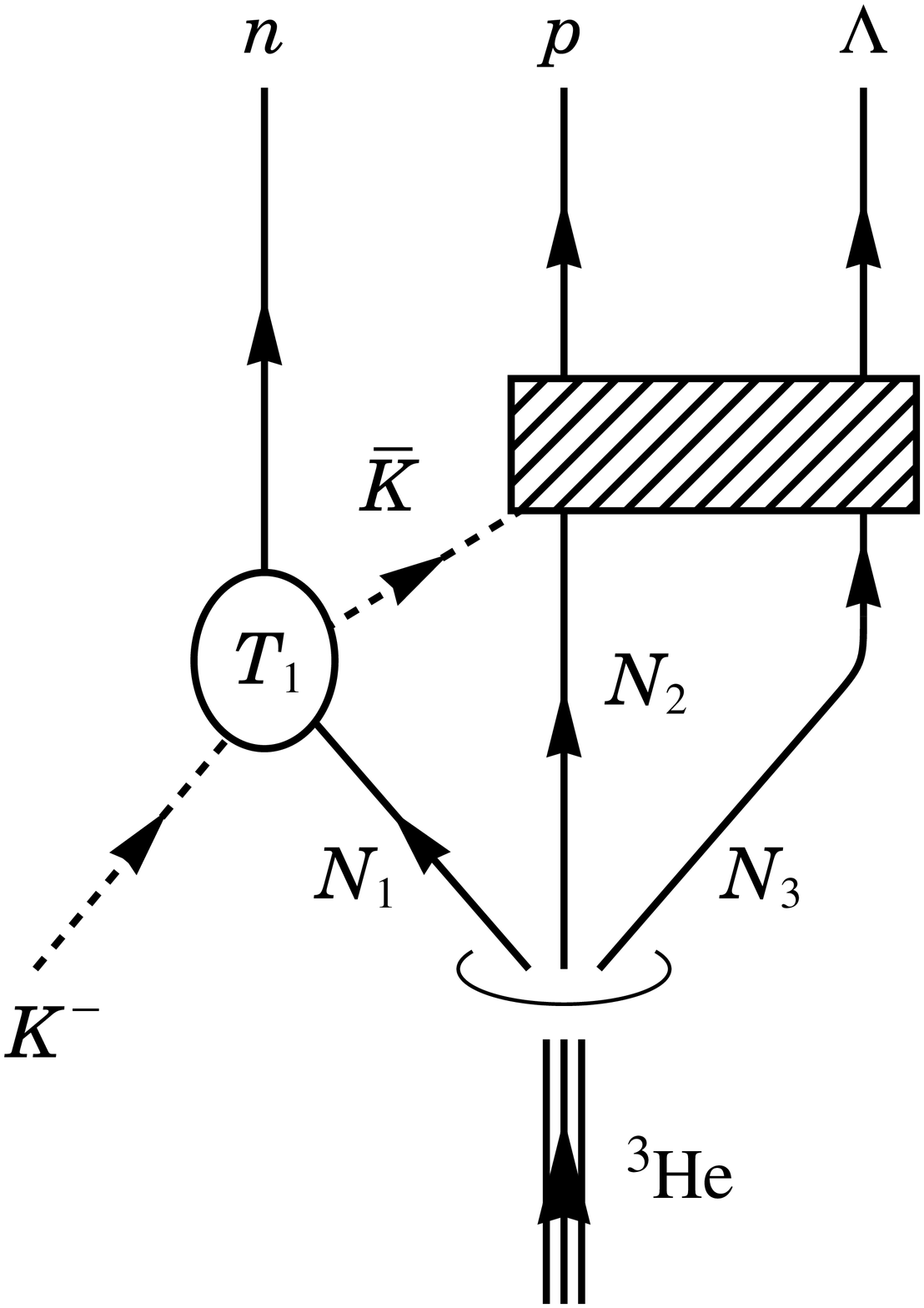}
    &  
    \includegraphics[scale=0.16]{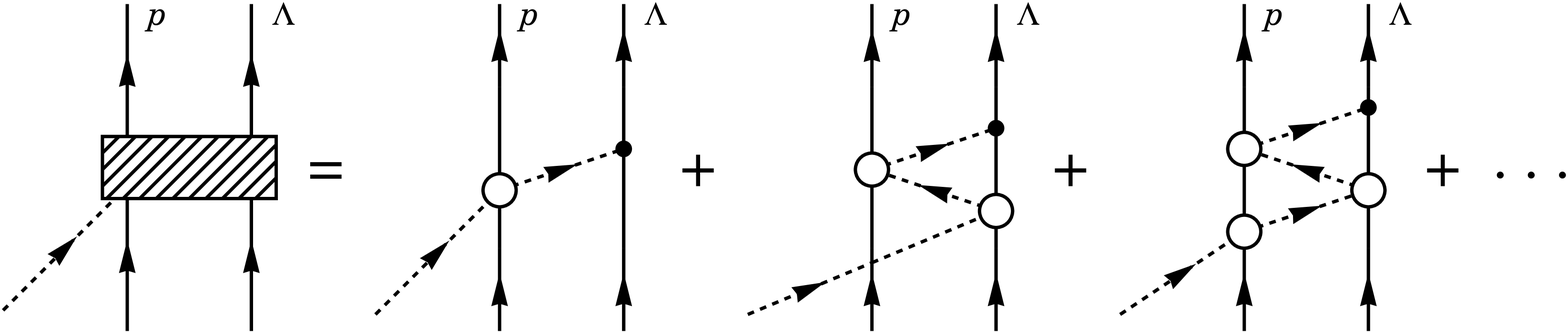}
    \\
    (a) & (b)
  \end{tabular}
  \caption{(a) Feynman diagram most relevant to the three-nucleon
    absorption of an in-flight $K^{-}$, and (b) multiple $\bar{K}$
    scattering and absorption~\cite{Sekihara:2016vyd}.  In (b), the
    dashed lines and circles represent $\bar{K}$ and $\bar{K} N \to
    \bar{K} N$ amplitude, respectively.}
  \label{fig:diag}
\end{figure}

In the J-PARC E15 experiment, they bombarded a ${}^{3} \text{He}$
target with $K^{-}$s of momentum $1 \text{ GeV} / c$ and observed the
$K^{-} {}^{3} \text{He} \to \Lambda p n$ reaction with forward neutron
in the final state and no spectator nucleon.  This reaction can be
expressed as the diagram in Fig.~\ref{fig:diag}.

In the first step of the reaction, the $K^{-}$ kicks out a fast and forward
final-state neutron and loses its energy.  The amplitude of this first
collision is calculated so as to reproduce the experimental values of
the cross sections of $K^{-} n \to K^{-} n$ and $K^{-} p \to
\bar{K}^{0} n$.  Because both the $K^{-} n \to K^{-} n$ and $K^{-} p
\to \bar{K}^{0} n$ cross sections have their local or global minima
when the final-state neutron goes forward, the $K^{-} {}^{3} \text{He}
\to \Lambda p n$ reaction favors the forward neutron emission compared
to the middle-angle emission.

\begin{figure}[b]
  \centering
  \includegraphics[width=8.6cm]{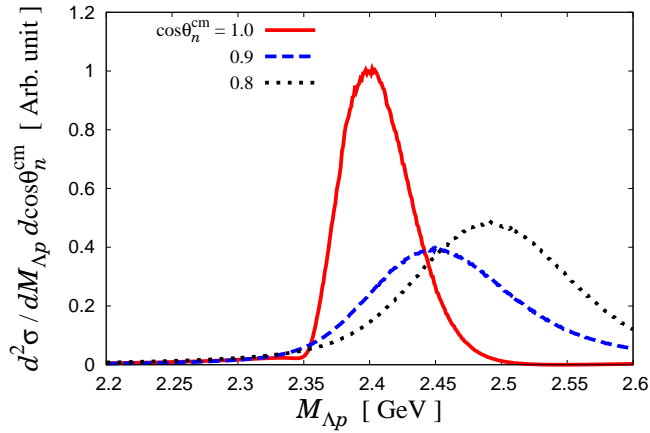}
  \caption{Differential cross section of the $K^{-} {}^{3}{\rm He} \to
    \Lambda p n$ reaction but neglecting $\bar{K} N N$ dynamics.}
  \label{fig:kin}
\end{figure}

Then, the slow $\bar{K}$ after the first collision propagates and is
absorbed into two nucleons from ${}^{3} \text{He}$.  An important
point is that this slow $\bar{K}$ can create a kinematic peak because
the propagating $\bar{K}$ can go almost on its mass shell, which
largely enhances the $\bar{K}$ propagator.  In order to see this
effect, we calculate the differential cross section of the $K^{-}
{}^{3} \text{He} \to \Lambda p n$ reaction according to the Feynman
diagram in Fig.~\ref{fig:diag} but neglecting the contribution from
the shaded box, \textit{i.e.}, making it unity.  This means that we
neglect dynamics of the slow $\bar{K}$ with two nucleons, which may
generate the $\bar{K} N N$ bound state.  The result is shown in
Fig.~\ref{fig:kin}.  As one can see, even if we do not have $\bar{K} N
N$ dynamics, we obtain a peak structure whose peak position is just
above the $\bar{K} N N$ threshold $= 2.37 \text{ GeV}$.  The
peak position shifts upward as the neutron angle becomes larger, which
can be explained by the kinematics of the quasi-elastic kaon
scattering in the first collision.

Now let us take into account $\bar{K} N N$ dynamics and transition
$\bar{K} N N \to \Lambda p$.  Because all the three particles,
$\bar{K}$ and two nucleons, are slow in the present reaction
mechanism, the multiple $\bar{K}$ scattering as in
Fig.~\ref{fig:diag}(b) should be essential in general.  Actually, if
we truncate the scattering in Fig.~\ref{fig:diag}(b) up to the first
term in the right-hand side [uncorrelated $\Lambda (1405) p$
  scenario], we cannot reproduce the behavior of the lower tail $\sim
2.3 \text{ GeV}$ in the experimental $\Lambda p$ invariant-mass
spectrum~\cite{Sekihara:2016vyd}.  In this study, we calculate the
multiple $\bar{K}$ scattering in the so-called fixed center
approximation~\cite{Bayar:2011qj, Bayar:2012hn}.  By including the
two-nucleon absorption width for $\bar{K}$ in a phenomenological way,
we obtain a $\bar{K} N N$ bound state with its pole position at $2354
- 36 i \text{ MeV}$~\cite{Sekihara:2016vyd}.  This multiple $\bar{K}$
scattering creates the peak structure in the $\Lambda p$
invariant-mass spectrum as the red thick line in Fig.~\ref{fig:dsdM}.
Our mass spectrum is consistent with the experimental one within the
present error.  An interesting finding is that we predict two peaks
across the $\bar{K} N N$ threshold in the spectrum.  The lower peak
comes from the $\bar{K} N N$ bound state, which reproduces the tail at
the lower energy $\sim 2.3 \text{ GeV}$ qualitatively well.  This
means that our spectrum supports the explanation that the E15 signal
in the ${}^{3} \text{He} ( K^{-} , \, \Lambda p) n$ reaction is indeed
a signal of the $\bar{K} N N$ bound state.  On the other hand, the
higher peak originates from the kinematics, \textit{i.e.}, from the
almost on-shell $\bar{K}$ denoted in Fig.~\ref{fig:kin}.  Because
using almost on-shell $\bar{K}$ is essential to make a $\bar{K} N N$
bound state in this reaction, this inevitably brings a kinematic peak
above the $\bar{K} N N$ threshold in the physical mass spectrum.

\section{Summary}

We expect that kaonic nuclei should exist owing to the strongly
attractive interaction between antikaon and nucleon.  Even the
existence of kaonic nuclei is still controversial, but a peak
structure which could be a signal of the simplest kaonic nucleus, the
$\bar{K} N N$ bound state, was recently found in the in-flight ${}^{3}
\text{He} ( K^{-} , \, \Lambda p ) n$ reaction in the J-PARC E15
experiment.

In order to understand the mechanism of the reaction, we theoretically
analyzed the reaction observed in the J-PARC E15 experiment.  We found
that, by detecting a fast and forward neutron in the final-state, an
almost on-shell $\bar{K}$ is guaranteed, which is essential to make a
bound state with two nucleons from ${}^{3} \text{He}$.  This almost
on-shell $\bar{K}$ can bring a signal of the $\bar{K} N N$ bound state
in the $\Lambda p$ invariant-mass spectrum, although it inevitably
brings a kinematic peak above the $\bar{K} N N$ threshold as well.  As
a consequence, we predicted two peaks across the $\bar{K} N N$
threshold in the spectrum: the lower peak coming from the $\bar{K} N
N$ bound state, and the higher one originating from the kinematics.

We finally note that the predicted two-peak structure is indeed
implied by the data of the E15 second run, where 30 times more
statistics of the same reaction are accumulated~\cite{Iwasaki:2017}.
This will support more strongly that the E15 signal in the ${}^{3}
\text{He} ( K^{-} , \, \Lambda p) n$ reaction is indeed a signal of
the $\bar{K} N N$ bound state.

\end{document}